\documentclass[twocolumn]{jpsj2} 
\usepackage{color}

\def\simge{\lower0.7ex\hbox{$\ \overset{>}{\sim}\ $}}
\def\simle{\lower0.7ex\hbox{$\ \overset{<}{\sim}\ $}}

\title{Continuous-time
Quantum Monte Carlo Approach for Impurity Anderson Models with
Phonon-assisted Hybridizations}

\author{Kazumasa HATTORI}

\inst{Institute for Solid State Physics, University of Tokyo, 5-1-5, Kashiwanoha, Kashiwa Chiba 277-8581, Japan
}

\abst{
We develop a continuous-time quantum Monte Carlo method based on a strong-coupling expansion for Anderson
impurity models with phonon-assisted hybridizations for arbitrary number
of phonon modes. As a benchmark, we investigate the two-channel Anderson
model with a single phonon, and numerically demonstrate that an 
SO(5) susceptibility composed of localized-electron charge and 
phonon-parity operators diverges logarithmically 
at non-Fermi liquid 
critical points in the model, which verifies the predictions by the
boundary conformal field theory[K. Hattori: Phys. Rev. B {\bf 85} (2012)
214411]. }

\kword{multi-channel Kondo effects, continuous-time quantum Monte Carlo,
non-Fermi liquid}

\begin{document}
\maketitle

\section{Introduction}

Kondo effects\cite{Kondo} in electron-phonon coupled systems have been attracted
great attention in recent years. Rare-earth based 
filled-skutterudites\cite{Sanada} and so-called 1-2-20
compounds\cite{1-2-20} are candidates for various Kondo effects due
to both magnetic and nonmagnetic origins.
There, well-localized f-electrons are located at a ``vibrating'' ion
inside a cage structure that provides conduction electrons to their Fermi surfaces.
For long time, systems with strong electron-phonon interactions 
have been considered as those exhibiting various types of the Kondo effects.\cite{Cox}
Recently, magnetically robust heavy-fermion states in the filled-skutterudite
SmOs$_4$Sb$_{12}$\cite{Sanada} have been attracted much attention
 due to the possible non-magnetic origin for the heavy
 fermion.\cite{Hat4level, Hotta}

Apart from complexities in the f-electron orbital degrees of freedom
 in these compounds, a prototype model
had been proposed already about thirty years ago by Yu and Anderson\cite{Yu}. They analyzed
 a local atomic oscillation coupled with spinless two-channel conduction
 electrons. The atom is assumed to oscillate along one direction, say z-axis,
 and thus, an electron-phonon coupling induces hybridizations between
conduction electrons with isotropic spherical wave and $p_z$-wave
 components. This is a so-called phonon-assisted hybridization process. 
Similar models have been analyzed in a line of discussions about possibility
 of two-channel Kondo effects in multi-level
 systems\cite{Vladar1,Vladar2,Vladar3,Fisher1,Fisher2, Kusunose}.

Several authors extended the model proposed by Yu and Anderson
to that includes
the spin degrees of freedom and the Coulomb interaction $U$ between localized
electrons with the different spins and analyzed it by using the Wilson's 
numerical renormalization group (NRG) method\cite{Dagott,Yashiki1,Yashiki2,Hotta2}. They found a line of 
non-Fermi liquid (NFL) fixed points characterized by spectra
realized in the magnetic two-channel Kondo model\cite{Cox} in the ground state phase diagram.
Recently, on the basis of the boundary conformal field theory (BCFT), 
we investigated the NFL and showed that the NFL in the
weak-coupling regime is qualitatively different from that in the conventional magnetic 
two-channel Kondo model. We also showed that  a  
crossover between the NFL of the magnetic two-channel Kondo model and 
the NFL in the weak-coupling regime, where SO(5) fluctuations---combined local-electron charge and
phonon parity fluctuations---are important, occurs, which can successfully explain the
NRG results\cite{Hat}.

A main purpose of this paper 
 is to develop  a numerical tool applicable to electron-phonon
systems with multi degrees of freedom, since, in systems with more than
one or two phonon modes, the Hilbert space becomes too large to be handled by, 
for example, the NRG or the exact diagonalization. 
For the Holstein phonon coupled with an electron density, 
an efficient quantum Monte Carlo method 
was proposed\cite{Werner2}. However, the technique there is not
 applicable to the model with phonon-assisted hybridization.  
In this paper, we develop a continuous-time quantum Monte Carlo (CTQMC) method\cite{Rubtsov, Werner1,Otsuki,Gull} to
 the Anderson impurity model with phonon-assisted
 hybridizations\cite{Dagott,Yashiki1} for multi phonon modes.


This paper is organized as follows. In Sect. \ref{sec-CTQMC}, we will show
a CTQMC algorithm for 
Anderson models with multi-channel conduction electrons and phonons. 
Benchmark tests in a small-size cluster problem will
be shown to convince readers of the efficiency of the method. 
We will apply this to the model with 
one-dimensional phonons\cite{Dagott,Yashiki1} and discuss the 
criticality of the model in Sect. \ref{MultiYAmodel}. Finally, Sect. \ref{Dis}
will summarize the present results and discuss possibilities of
 application of the present method to more complicated systems.

\section{Continuous-time Quantum Monte Carlo Method}\label{sec-CTQMC}
{In this section, we will present our CTQMC algorithm for 
Anderson-type models with phonon-assisted hybridizations.
For details of the basic algorithm for the impurity Anderson model, 
see the review paper.\cite{Gull} }
After presenting models we use in Sect. \ref{Model}, we will explain our
algorithm of CTQMC in Sect. \ref{algorithm} and then show a benchmark
result for  a three-site cluster model in Sect. \ref{bench}. 
\subsection{Model}\label{Model}
We investigate an impurity Anderson model with phonon-assisted
hybridization\cite{Dagott,Yashiki1} generalized to one with
$M$ phonon modes, 
\begin{eqnarray}
H&=&H_c+H_l+H_{ph}+V+V^{\dagger},\label{H}\\
H_c&=&\sum_{\alpha=0}^{M}\sum_{\sigma}\int dk\ \epsilon_{k\alpha}c^{\dagger}_{k\alpha\sigma}c_{k\alpha\sigma},\\
H_l&=&\sum_{\sigma}\epsilon_{\rm f} f^{\dagger}_{\sigma}f_{\sigma}
+Uf^{\dagger}_{\uparrow}f_{\uparrow}f^{\dagger}_{\downarrow}f_{\downarrow},\\
V&=&\sum_{\alpha=0}^M\sum_{\sigma}V_{\alpha\sigma},\\
V_{\alpha\sigma}&=&\frac{v_{\alpha\sigma}}{\sqrt{M}}X_{\alpha}
f^{\dagger}_{\sigma} \int dk \ c_{k\alpha\sigma}.\label{Hv}
\end{eqnarray}
Here, the conduction electrons are written in the bases of spherical
wave and $c_{k\alpha\sigma}$ indicates the conduction electron creation operator
with the radial wavenumber $k$, the orbital $\alpha=0,1,\cdots$, or $M$, and the spin
$\sigma=\uparrow$ or $\downarrow$. $f^{\dagger}_{\sigma}$ represents the
localized
electron creation operator with the spin $\sigma$ and we assume it is isotropic $s$-orbital. 
$\epsilon_{\rm f}$ and $U$ are the
localized electron energy level and the Coulomb
interaction, respectively. $v_{\alpha\sigma}$ represents hybridization between localized and
conduction electrons with $\alpha$
orbital and $\sigma$ spin. $X_{\alpha}=X_{\alpha}^{\dagger}$ indicates 
phonon-displacement operators that are dimensionless quantities scaled by an appropriate length scale
 and have the same symmetry as the orbital $\alpha$ to make the
 Hamiltonian invariant. 
$H_{ph}$ represents the Hamiltonian for phonons. 
In order to make computational cost small, we restrict ourselves to considering $H_{ph}$ in which 
each $X_\alpha$ does not couple.\cite{well} For simplicity, throughout this paper, we
will use a harmonic oscillator model for $H_{ph}$.

Hybridization processes without phonon assists are included in the $V$
 term with $\alpha=0$ in eq. (\ref{Hv}); $X_{0}\equiv 1$. 
Corresponding to this, $c_{k0\sigma}^{\dagger}$ represents
 the creation operator of an $s$-orbital electron.
When phonon oscillation amplitudes are small, $X_{\alpha}$'s
($\alpha=1,2, \cdots, M$) are, in the
first-order approximation, represented by the linear-displacement
operators $x_{\alpha}$'s, which couple with $p$-wave components of
conduction electrons $\int dk c_{k\alpha\sigma}$ with
 $\alpha=1,2,\cdots, M$ around the impurity site in eq. (\ref{Hv}). 
Here, in this case, $M$
 represents the dimensionality of the oscillation mode.
One can also construct
models that include processes with higher-order displacements such
as $x_{\beta}x_{\gamma}$, $x_{\beta}x_{\gamma}x_{\delta}$, and so
on, which couple with higher-order harmonics of spherical bases for
conduction electrons.\cite{note}
 Although we do not discuss such models in this paper, our 
CTQMC method can handle these general hybridization processes.

\subsection{Algorithm}\label{algorithm}
In this subsection, we summarize the algorithm\cite{issp} of the CTQMC
 applied to the model (\ref{H}) on the basis of
strong coupling expansion\cite{Werner1,Gull,Werner3,Haul}, {\it i.e.,} perturbative expansions of $V$. 

In terms of the infinite series of $V$ and $V^{\dagger}$, 
the partition function $Z$ for the model (\ref{H}) is expressed as
\begin{eqnarray}
Z\!=\!Z_cZ_lZ_{ph}\Bigg\langle T_{\tau}\exp\Bigg\{\!\!-\int_0^{\beta}\!\!d\tau
 \Big[V(\tau)+V^{\dagger}(\tau)\Big]\Bigg\}\Bigg\rangle_0, \label{Z}
\end{eqnarray}
where $T_{\tau}$ represents time-ordered product and $\beta=1/T$ with
$T$ being temperature. $Z_c$, $Z_l$ and $Z_{ph}$ are the partition function of
non-interacting conduction electrons, that for localized electrons, and
that for local phonons, respectively, and 
\begin{eqnarray}
\langle A\rangle_0\equiv \frac{{\rm Tr}\{A
e^{-\beta(H_c+H_l+H_{ph})}\}}{Z_cZ_lZ_{ph}}. 
\end{eqnarray}

As discussed by Werner
{\it et al.},\cite{Werner1} 
eq. (\ref{Z}) is evaluated by Monte Carlo simulations, in which the positions of $V(\tau)$ and
 $V^{\dagger}(\tau)$ 
 along the imaginary time $\tau$ and also the perturbation order are sampled. 
In addition to the conduction- and the local-electron parts in $Z$, we need to calculate a part due to the phonons.
For this, we need to evaluate 
\begin{eqnarray}
 \langle T_{\tau}x_{\alpha}({\tau_1})x_{\alpha}({\tau_2})\cdots
  x_{\alpha}(\tau_{2k_{\alpha}})\rangle_{ph}. \label{xxx}
\end{eqnarray}
Here, $\langle A\rangle_{ph}\equiv {\rm Tr} \{ Ae^{-\beta H_{ph}}
\}/Z_{ph}$, and $k_{\alpha}$ is an integer with $2k_{\alpha}$ being the
perturbation order of the $\alpha$th
phonon-assisted term, {\it i.e.}, the total number of vertices
$v_{\alpha\uparrow}$ and $v_{\alpha\downarrow}$ in a configuration
considered. 
Note that different
$\alpha$'s do not couple, since we have assumed that in $H_{ph}$ each of
the phonon mode is
decoupled.

 Unlike the case of Holstein phonons\cite{Werner2}, a simple
canonical transformation does not work on absorbing the phonon terms into 
phase factors.
{ This originates in the facts that the model is one with multiorbital
in general and the off-diagonal hybridization density couples with the
phonons, while there is the Coulomb interaction only for the $f$ electrons, in contrast to the case in the Holstein-Hubbard model\cite{Werner2} where 
the density does in the electron-phonon coupling term. 
}
For actual calculations of eq. (\ref{xxx}), we introduce a cutoff for the phonon Hilbert
space for each $\alpha$: $N_{\rm cut}$. This part might not be a smart
way, but it is at least efficient when one investigates models with
multi-phonon modes and multi-orbital conduction electrons.

In practice, to make the computations fast, we store intermediate matrices
in the matrix product calculations of eq. (\ref{xxx}) and re-use
them at later steps in the Monte Carlo simulations\cite{Haul} and also
use the fast-update algorithm\cite{Rubtsov}.

\begin{figure}[t!]
\begin{center}
    \includegraphics[width=0.47\textwidth]{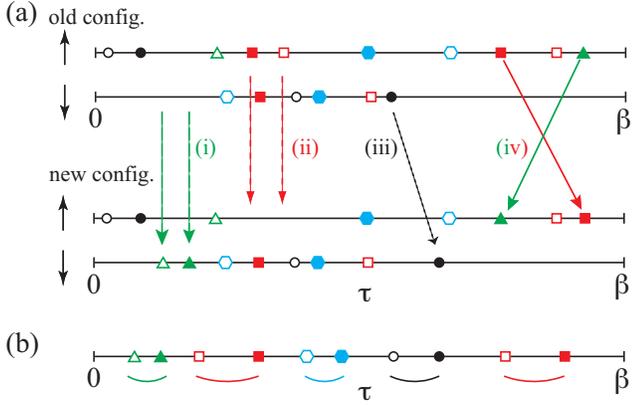}
\end{center}
\vspace{-2mm}
\caption{(Color online) (a) Four update processes. The horizontal
 line represents imaginary time axis and the upper two represent ``old''
 configuration with the spin $\uparrow$ and $\downarrow$, and the lower two do ``new'' one. 
Open (filled) symbols represents $V_{\alpha\sigma}$
 ($V^{\dagger}_{\alpha\sigma}$) and the types of symbols identify the
 orbital index $\alpha$. 
Note that every open (filled) symbol is
 sandwiched by two filled (open) symbols and the total number of the open and
 the filled symbols are the same. 
(i) Inserting two
 vertices, (ii) removal of two vertices, (iii) shifting a vertex, and
 (iv) exchanging two vertices. (b) Typical vertex configuration without
 the exchange update. }
\label{fig-update}
\vspace{-5mm}
\end{figure}

For carrying out Monte Carlo samplings in the whole phase space of the partition function $Z$,
we need to introduce appropriate update operations.
In a single-impurity Anderson model, conventional updates are known to
be\cite{Werner1} (i) inserting two vertices
$|v_{\alpha\sigma}|^2$, (ii) removing them, and (iii) shifting a vertex position in
the imaginary time, as shown in Fig. \ref{fig-update} (a). 
A new update operation is necessary for realizing the random
walk satisfying the Ergodicity in the present model in addition to
conventional ones. That is (iv) exchanging two
vertices $V_{\alpha\sigma}$ and $V_{\beta\sigma}$, or $V_{\alpha\sigma}^{\dagger}$
and $V^{\dagger}_{\beta\sigma}$ with $\alpha\ne \beta$,  as depicted
in Fig. \ref{fig-update} (a). Without this update, the vertex sequence along
the imaginary-time axis is always paired in the same $\alpha$, 
 which is
only a part of the whole phase space. See Fig. \ref{fig-update} (b). 
Upon the exchange update, only the conduction and phonon parts are
affected, while the local-electron part is unchanged.

\subsection{A benchmark}\label{bench}
To check the algorithm explained in Sect. \ref{algorithm}, in this
subsection, we will show results
for a 
finite-size system and compare the
results by CTQMC with those by the exact diagonalization. 

Here, we
consider 
a harmonic-oscillator model for $H_{ph}$ with $M=1$
 as an illustration of the efficiency of our method. We use a three-site
 model, which is equivalent to replacing the
conduction electrons $c_{k0\sigma}$ by one electron $c_{\sigma}$ and
$c_{k1\sigma}$ by $p_{\sigma}$. 
Hamiltonian for this system is given as, 
\begin{eqnarray}
H_{\rm 3
 sites}&=&\sum_{\sigma}\Big[(v_0f^{\dagger}_{\sigma}c_{\sigma}+v_1 x f^{\dagger}_{\sigma}p_{\sigma}+{\rm
 h.c.})+\epsilon_{\rm
 f}f^{\dagger}_{\sigma}f_{\sigma}\Big]\nonumber\\
&+&Uf^{\dagger}_{\uparrow}f_{\uparrow}f_{\downarrow}^{\dagger}f_{\downarrow}
+\Omega \Big(b^{\dagger}b+\frac{1}{2}\Big),\label{Hbench}
\end{eqnarray}
where $x\equiv b^{\dagger}+b$ with $b^{\dagger}$ being the phonon
creation operator  and $\Omega$
is the phonon energy. $U$ is set to the energy unit $U=+1$, and we use $v_0/U=0.2$, $v_1/U=0.18$,
$\epsilon_{\rm f}/U=-0.5$, and $\Omega/U=0.2$. In the
following, we use the same $N_{\rm cut}$
both for the CTQMC and the exact diagonalization. Thus, the two results
should be the same within the statistical errors in the CTQMC.

\begin{figure}[t]

\begin{center}
    \includegraphics[width=0.46\textwidth]{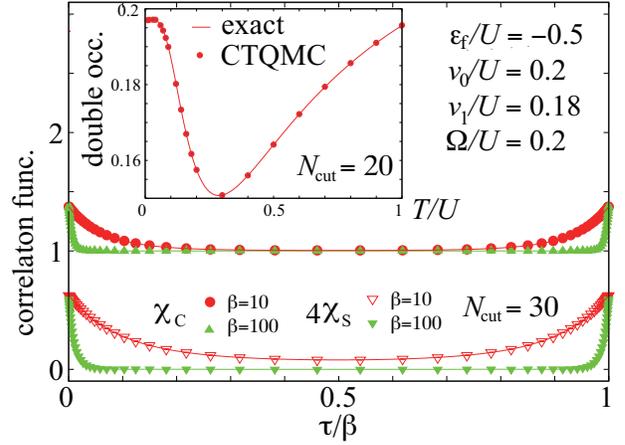}
\end{center}
\vspace{-3.5mm}
\caption{(Color online) Imaginary time dependence of charge and spin susceptibilities
 for $\beta=10$ and $100$. Symbols represent the CTQMC results
 and lines are those of the exact diagonalization. $N_{\rm cut}=30$ and 
 $\epsilon_{\rm f}/U=-0.5$, $v_0/U=0.2$, $v_1/U=0.18$, and $\Omega/U=0.2$. 
Inset: temperature dependence of double occupancy for
 $N_{\rm cut}=20$ and the same parameters as in the main panel. Error bars are smaller than the
 symbol size.}
\label{fig-bench}
\vspace{-2mm}
\end{figure}

Figure \ref{fig-bench} shows the imaginary time dependence of the charge
 susceptibility:
\begin{eqnarray}
\chi_c(\tau)=\langle T_{\tau}
[n_{\uparrow}(\tau)+n_{\downarrow}(\tau)][n_{\uparrow}(0)+n_{\downarrow}(0)]\rangle,
\end{eqnarray}
and the spin susceptibility: 
\begin{eqnarray}
\chi_s(\tau)=\langle T_{\tau}
[n_{\uparrow}(\tau)-n_{\downarrow}(\tau)][n_{\uparrow}(0)-n_{\downarrow}(0)]\rangle/4,\label{chis}
\end{eqnarray}
for $\beta=10$ and $100$ with $N_{\rm
cut}=30$. Inset of Fig. \ref{fig-bench} shows 
temperature dependence of double
occupancy $\langle n_{\uparrow}n_{\downarrow}\rangle$ for 
$N_{\rm cut}=20$,
where $n_{\sigma}=f_{\sigma}^{\dagger}f_{\sigma}$ and 
$\langle A\rangle$ represents the thermal average of operator $A$. The statistical
errors for CTQMC data are smaller than the symbol sizes. 
One can clearly see that the CTQMC data reproduce the exact diagonalization
ones within the statistical errors, which confirms the efficiency of our method.

\section{Analysis of $M=1$ Harmonic Model}\label{MultiYAmodel}
In this section, we will show numerical results of the model with $M=1$ and $H_{ph}=\Omega (b^{\dagger}b+1/2)$ as in
eq. (\ref{Hbench}) and we use spin-independent hybridizations in
eq. (\ref{Hv}): $v_{0\sigma}=v_0$ and $v_{1\sigma}=v_1$ as in Sect. \ref{bench}. 
We use $\epsilon_{k\alpha}=v_F(k-k_F)$ with $v_F$($k_F$) being the
Fermi velocity (wavenumber) for all $\alpha$. Band width $2D$ is set to
$D\equiv v_F \Lambda$, where $\Lambda$ is the
cutoff for the wavenumber and the density of states are set to a
constant $1/(2D)$ from $-D$ to $D$ by choosing appropriate values of
$v_F$ and $\Lambda$. Throughout this section, the cutoff of the phonon
Hilbert space is set to $N_{\rm cut}=20$. 

In previous studies of this model, a line of 
two-channel Kondo like NFL fixed points is found for $U\ne 0$\cite{Yashiki1,Yashiki2}.
Based on the NRG and the BCFT,\cite{Hat} the NFL for small-$U$ region turns out
to be
qualitatively different from that in the magnetic two-channel Kondo model. In
particular, SO(5) symmetric operators were identified and it was predicted
that susceptibilities 
of five-dimensional vector operators in the SO(5) sector diverge logarithmically at low
temperatures. In the following, we concentrate on examining the
divergence of the susceptibilities at the critical points of this model.  

\subsection{Susceptibilities}\label{chi}
For the discussions in Sect. \ref{numericalresult}, we introduce following
three susceptibilities, which are expected to show singular temperature
dependence at the critical points. 

First, we define a spin
susceptibility given by
\begin{eqnarray}
\chi_s^{z}(T)=\int_0^{\beta}d\tau\chi_s(\tau),
\end{eqnarray}
where $\chi_s(\tau)$ is given by eq. (\ref{chis}).
The second is a coupled localized-electron's charge and phonon-parity 
susceptibility, which corresponds to an SO(5) vector
susceptibility\cite{Hat} with slight (not essential) simplifications,
\begin{eqnarray}
\chi^{z}_{{\mathcal P}p}(T)&=&\int_0^{\beta}d\tau\Big[\langle
 T_{\tau} {\mathcal P}(\tau)p_z(\tau) {\mathcal P}(0)p_{z}(0)
 \rangle \nonumber\\
&&-\langle {\mathcal P}(0)p_z(0)
\rangle^2\Big],\\
p_z&\equiv& \sum_{n=0}^{N_{\rm cut}} (-1)^n|n\rangle\langle n|,\\
{\mathcal P}&\equiv& (n_{\rm f}-1)^2, 
\end{eqnarray}
where $n$ represents the phonon number and $n_{\rm f}=\sum_{\sigma}f^{\dagger}_{\sigma}f_{\sigma}$.
Note that ${\mathcal P}$ is the projection operator on $n_{\rm f}=0$ and
2 subspaces.
The third one is a parity fluctuation of the phonons written as 
\begin{eqnarray}
\chi_{p}^{z}(T)\!\!\!\!\!\!&=&\!\!\!\!\!\!\int_0^{\beta}\!\!\!d\tau
\Big[
\langle T_{\tau} p_z(\tau) p_{z}(0) \rangle
-\langle p_z(0)\rangle^2
\Big].
\end{eqnarray}

According to the BCFT,\cite{Hat} 
$\chi_s^z(T)$ and $\chi_{{\mathcal P}p}^z(T)$ diverge logarithmically
$\sim -\ln T$ at the critical
points of this model. 
It has been demonstrated that the diverging parts in $\chi_{p}^z(T)$ arise from 
 $\chi_{{\mathcal P}p}^z(T)$ and the important parts for the divergence
 originate in the coupled localized-electron charge sectors with $n_{\rm f}-1=\pm 1$ and the
 parity fluctuations.

\begin{figure}[t!]

\begin{center}
    \includegraphics[width=0.46\textwidth]{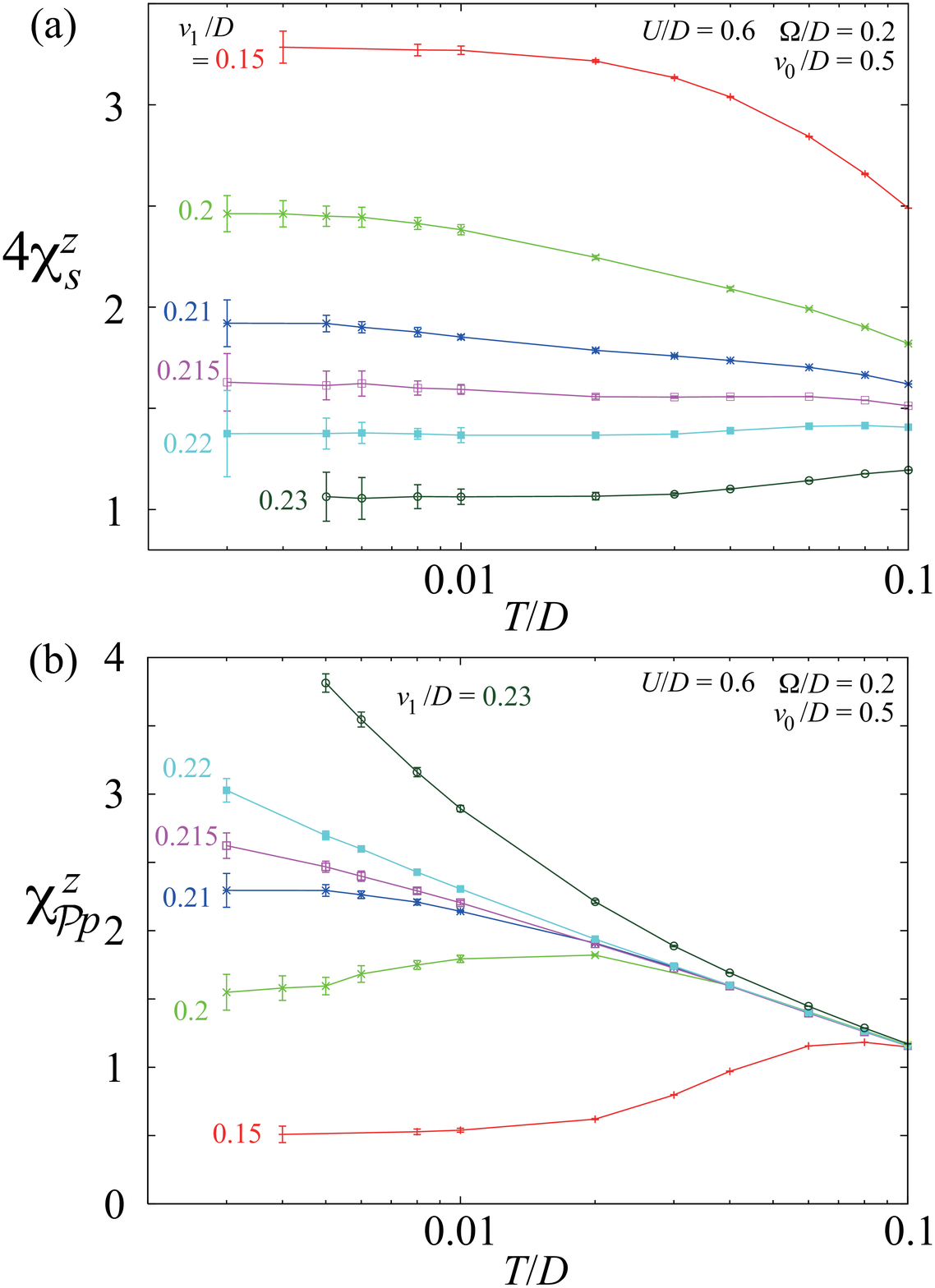}
\end{center}
\vspace{-3mm}
\caption{(Color online) Temperature dependence of (a) $\chi_s^z(T)$ and (b)
 $\chi_{{\mathcal P}p}^z(T)$ for several values
 of $v_1$'s, $U/D=0.6$,
 $v_0/D=0.5$, and $\Omega/D=0.2$ with $N_c=20$. }
\label{fig-3}

\end{figure}
\subsection{Numerical results}\label{numericalresult}
{
Before discussing numerical results, we first comment about some 
technical aspects. First, we have found no noticeable negative sign problem
in our line of calculations in this paper
as in the calculations for the Anderson model.\cite{Werner2} 
Secondly, we have checked that the cutoff $N_{\rm cut}=20$ is sufficiently large for all
temperature range we have examined. This can be checked by calculating 
probability distribution of phonon number $n$ in the CTQMC; the
probability for $n=N_{\rm cut}$ is zero throughout the CTQMC sampling.
}

Now, let us start to discuss the results for small $U$ regime, $U/D=0.6$. 
Figure \ref{fig-3} shows temperature dependence of the
susceptibilities $\chi_s^z(T)$ and $\chi_{{\mathcal P}p}^z(T)$. The spin susceptibility $\chi_s^z(T)$
[Fig. \ref{fig-3} (a)] shows
no noticeable temperature dependence at low temperatures, while $\chi_{{\mathcal
P}p}^z(T)$ shows logarithmic divergence for $v_1/D=0.22$ as shown in Fig. \ref{fig-3} (b).
The logarithmic divergence is expected to appear at the critical point, 
and thus, the critical point is located near $v_1/D\sim 0.22$. 
For a
putative logarithmic singularity in $\chi_s^z(T)$, we cannot find
noticeable one. We consider that the absolute value of the
singularity is so small that it cannot be detectable within the present
error bars in this small $U/D=0.6$.

\begin{figure}[t!]

\begin{center}
    \includegraphics[width=0.46\textwidth]{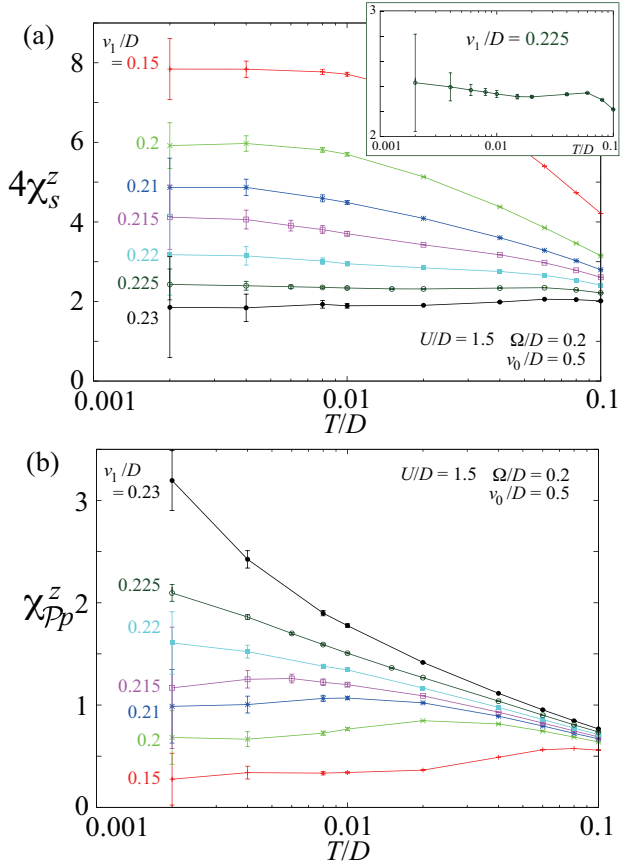}
\end{center}
\vspace{-3mm}
\caption{(Color online) Temperature dependence of (a) $\chi_s^z(T)$ and (b)
 $\chi_{{\mathcal P}p}^z(T)$ for several values
 of $v_1$'s, $U/D=1.5$,
 $v_0/D=0.5$, and $\Omega/D=0.2$ with $N_c=20$. Inset in (a): zoom-up
 for $v_1/D=0.225$.}
\label{fig-4}
\vspace{-3mm}
\end{figure}

For larger $U$, the spin susceptibility shows a 
logarithmic increase at the critical point. Figure \ref{fig-4} shows
$\chi_s^z(T)$ and $\chi_{{\mathcal P}p}^z(T)$
 as a function of $T$ for $U/D=1.5$. One can see that 
$\chi_s^z(T)$ [Fig. \ref{fig-4} (a)]
and $\chi_{{\mathcal P}p}^z(T)$ [Fig. \ref{fig-4} (b)] show logarithmic
increases  near the critical point $v_1/D\sim 0.225$. See the zoom up
for $\chi_s^z(T)$ for $v_1/D=0.225$.
This indicates that the singularity in the spin sector becomes more
prominent for $U/D=1.5$ than for the smaller $U/D=0.6$ in
Fig. \ref{fig-3}.
As for $\chi_{{\mathcal
P}p}^z(T)$, the absolute value becomes smaller than that for $U/D=0.6$.
This is because the system approaches the local moment regime as $U$
increases.
This crossover is consistent with the results in the NRG
and the discussion in the BCFT.\cite{Hat}

For both $U/D=0.6$ and $1.5$, strong increases in $\chi_{{\mathcal
P}p}^z(T)$ for $v_1$ larger than the critical value are due to the existence
of nearly degenerate nonmagnetic states in the large $v_1$ limit, as discussed in the early 
work.\cite{Yashiki2} Thus, it is expected that they decrease for $T$
smaller than the gap, but it is known that this is very
small\cite{Yashiki2} and the CTQMC
cannot reach such a small temperature.

\begin{figure}[t!]

\begin{center}
    \includegraphics[width=0.46\textwidth]{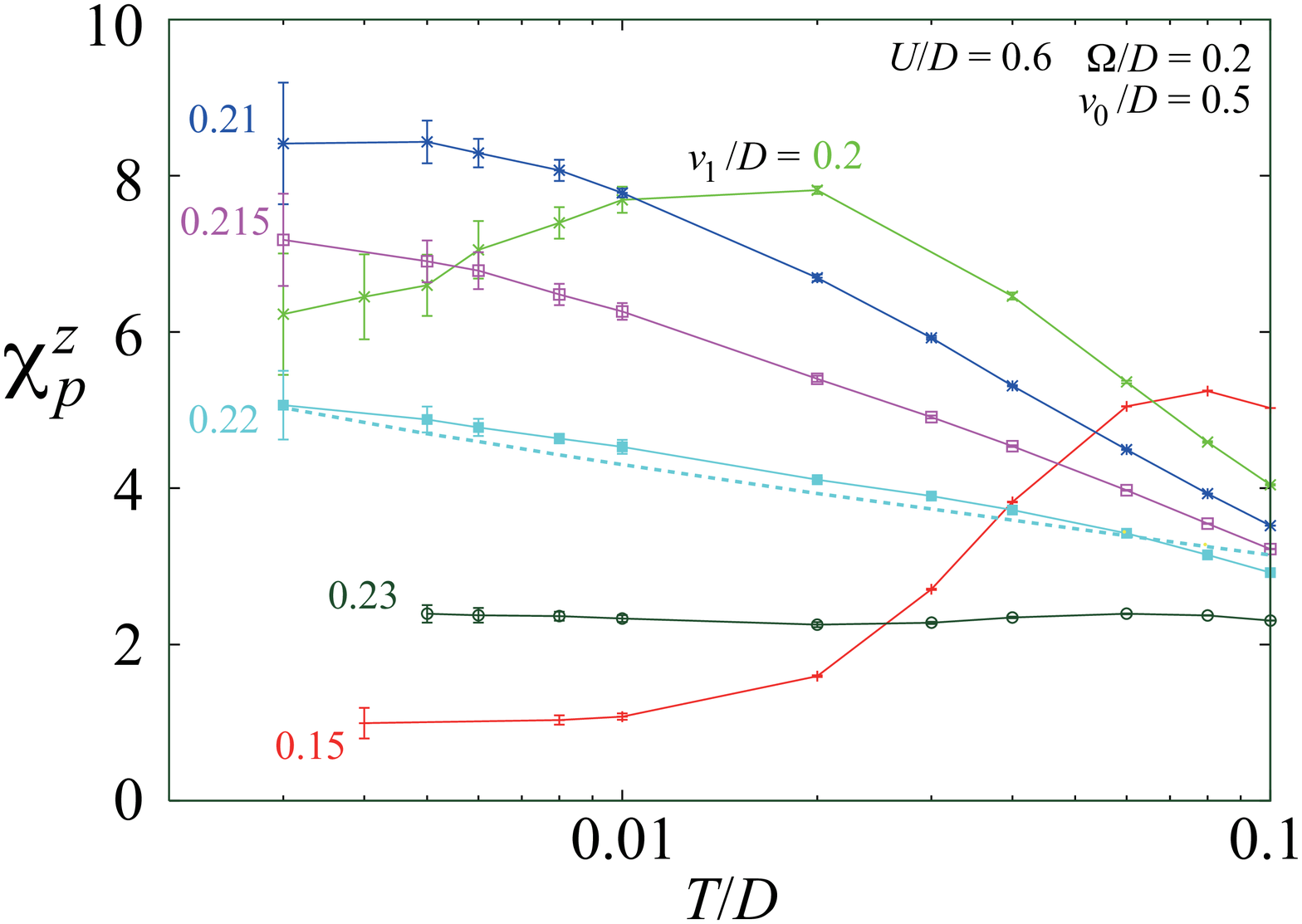}
\end{center}
\vspace{-3mm}
\caption{(Color online) Temperature dependence of $\chi_p^z(T)$ 
 for several values
 of $v_1$'s, $U/D=0.6$,
 $v_0/D=0.5$, and $\Omega/D=0.2$ with $N_c=20$. A dashed line represents
 $\chi_{{\mathcal P}p}^z(T)+2$ for $v_1/D=0.22$ from Fig. \ref{fig-3} (b).}
\label{fig-5}
\vspace{-2mm}
\end{figure}

We have also confirmed that the logarithmic temperature dependence in
$\chi_p^z(T)$ near the critical point comes only from the subspace 
projected by $\mathcal P$.
Figure \ref{fig-5} shows the temperature dependence of $\chi_p^z(T)$ for
$U/D=0.6$. For comparison, $\chi_{{\mathcal P}p}^z(T)$ near the critical
point ($v_1/D=0.22$) shifted by 2 is plotted by a dashed line. 
As seen in Fig. \ref{fig-5}, both $\chi_p^z(T)$ and
$\chi_{{\mathcal P}p}^z(T)$ show logarithmic increases for
$v_1/D=0.22$. Importantly, the logarithmic increase in $\chi_p^z(T)$ is 
quantitatively the same as those seen in $\chi_{{\mathcal P}p}^z(T)$
represented by the dashed line. Since $\chi_{{\mathcal P}p}^z(T)$ is a part
projected by $\mathcal P$ out of $\chi_p^z(T)$, this confirms that the
singularity originates in the sector projected by $\mathcal P$, {\it
i.e.}, $n_{\rm f}=0$ and $2$ sectors, which is consistent with the
prediction by the BCFT.\cite{Hat}

\section{Discussions and Summary}\label{Dis}
We have developed a continuous-time quantum Monte Carlo method for
impurity Anderson models with phonon-assisted hybridizations.
The method can be applicable to models with several phonon modes and
also non-harmonic phonon models within the restriction that each of
phonon modes is decoupled. Even under this restriction, one can analyze
various interesting models, such as a model with three-dimensional anharmonic potential
$V(x,y,z)={\rm harmonic\ terms}+\eta_x x^4+\eta_y y^4+\eta_z z^4$ with
$\eta_i$ $(i=x,y$, or $z)$ being 
anharmonic parameters and one with an infinite-well potential as noted before.\cite{well} Advantage of
using the CTQMC to solve models with multi phonon degrees of freedom is
that one can treat its large Hilbert space in as small computational
cost as in a single-phonon case (partly double for the exchange update). 

An important point is that the computational
cost decreases as the number of phonon modes {\it increases}, since, 
compared with the single-phonon case, 
the perturbation order for each mode $2k_{\alpha}$ in eq. (\ref{xxx})  becomes
smaller in multi-phonon cases. This is because the total
perturbation order is not sensitive to the number of modes 
 as discussed in the CTQMC for SU($N$) Coqblin-Schrieffer
model.\cite{Otsuki2} 
{ Thus, the perturbation order per orbital decreases, which
leads to reduction of the number of the matrix products in
eq. (\ref{xxx}). This greatly 
 reduces  computational costs when the number of modes increases}
and opens possibilities for exploring various exotic Kondo effects in
systems with multidegrees of freedom, which have never been reached by existing
numerical (and also analytical) methods.

Additional Holstein phonons are easily handled by the canonical
transformation as was done by Werner and Millis.\cite{Werner2}
Extending the local electron part to one with orbital degrees of freedom is straight
forward with slightly increasing computational costs and this is
necessary extension for investigating more realistic systems. These are
our future problems.

In the final part in Sect. \ref{MultiYAmodel}, we have applied our CTQMC
algorithm to the two-channel Anderson model with
phonon-assisted hybridizations. 
The results have revealed that fluctuations for the coupled electron-phonon
degrees of freedom
diverge logarithmically at low temperatures near the critical point. This is consistent with the previous theoretical
analysis\cite{Hat} and demonstrates the validity of the present method.
As a next step, analyses of models with multi phonon modes
are now in progress.


\section*{Acknowledgment}
The author thanks H. Tsunetsugu and T. Sato for fruitful discussions. 
This work is supported by
KAKENHI (Grant No. 30456199) and by a Grant-in-Aid for 
Scientific Research on Innovative
Areas ``Heavy Electrons'' (Grant No. 23102707) of The Ministry of
Education, Culture, Sports, Science, and Technology, Japan.
A part of the numerical
calculations was done at the Supercomputer Center at ISSP,
University of Tokyo and also 
at Information Technology Center, University of Tokyo.


\begin{thebibliography}{99}
\bibitem{well} This includes, {\it e.g.}, an infinite-well potential defined
	by {$V(x,y,\cdots)$$=$$0$ for $|x|$$\le$$a$, $|y|$$\le$$b$, and $\cdots$,
	while $V(x,y,\cdots)$$=$$\infty$ for others, where $2a$, $2b$,
	$\cdots$ are the width of the well for $x$, $y$, $\cdots$ 
	directions, respectively.}
\bibitem{note} Of course, {operators in a same irreducible
	representation can couple with each other.} For example, isotropic combinations such as
	$\sum_{\beta}x_{\beta}^2$ couple to $s$-wave components of
	conduction electrons.
\end{thebibliography}
\end{document}